\documentclass[epsf,twocolumn,showpacs,preprintnumbers]{revtex4}

\usepackage{graphics}
\usepackage{graphicx}
\usepackage{dcolumn} 
\usepackage{bm}
\usepackage{color}
\usepackage{epsfig}
\usepackage{comment}
\newcommand{\bfpo}{BaFe$_2$(PO$_4$)$_2$}
\newcommand{\ch}{${\cal C}$}
\begin{document}
\title{
Tuning ferromagnetic BaFe$_2$(PO$_4$)$_2$ through a high Chern number
topological phase
}
\author{Young-Joon Song$^1$}
\author{Kyo-Hoon Ahn$^1$}
\author{Warren E. Pickett$^2$}
\email{pickett@physics.ucdavis.edu} 
\author{Kwan-Woo Lee$^{1,3}$}
\email{mckwan@korea.ac.kr}
\affiliation{
 $^1$Department of Applied Physics, Graduate School, Korea University, Sejong 33019, Korea\\
 $^2$Department of Physics, University of California, Davis, CA 95616, USA\\
 $^3$Department of Display and Semiconductor Physics, Korea University, Sejong 33019, Korea
}
\date{\today}
\begin{abstract}
There is strong interest in discovering or designing wide gap Chern insulators.
Here we follow a Chern insulator
to trivial Mott insulator transition versus interaction strength $U$ in a
honeycomb lattice Fe-based transition metal oxide, discovering that a
spin-orbit coupling  energy scale $\xi$=40 meV can produce and maintain a topologically
entangled Chern insulating state against large band structure changes arising from
an interaction strength $U$ up to 60 times as large. 
Within the Chern phase the minimum gap switches from the zone corner $K$ to the
zone center $\Gamma$ while maintaining the topological structure.
At a critical strength $U_c$,
the continuous evolution of the electronic structure encounters a gap closing then
reopening, upon which the system reverts to a trivial Mott insulating phase. 
This Chern insulator phase of honeycomb lattice 
Fe$^{2+}$ BaFe$_2$(PO$_4$)$_2$ corresponds to a large Chern number
\ch~= --3 that will provide enhanced anomalous
Hall conductivity due to the associated three edge states
threading through the bulk gap of 80 meV.

\end{abstract}
\maketitle

\section{Background}

\begin{figure}[tbp]
{\resizebox{8.0cm}{5.5cm}{\includegraphics{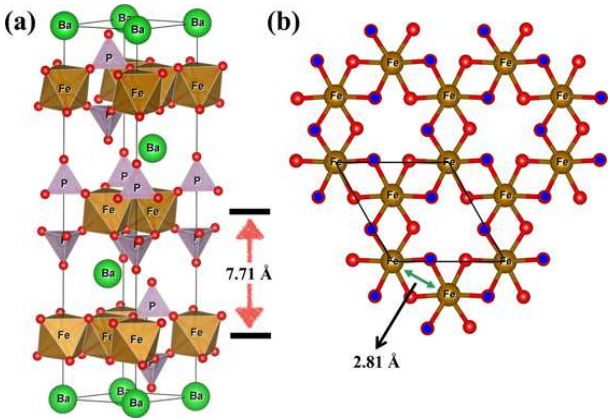}}}
\caption{(color online) The rhombohedral structure of \bfpo (left panel),
consisting of honeycomb sublattices of FeO$_6$ octahedra (right panel, top view) 
and intervening PO$_4$ tetrahedra and Ba$^{2+}$ ions.
The primitive cell (solid lines in the right panel) contains two Fe$^{2+}$ ions.
The inter- and intralayer Fe-Fe distances, indicating highly 2D character, 
are provided.
}
\label{structure}
\end{figure}

The quantum anomalous Hall (QAH) insulator, also known as the Chern insulator,
is a two-dimensional (2D) ferromagnetic (FM) insulator with a nonzero Chern number,
resulting in a quantized boundary anomalous Hall conductivity
without an external magnetic field 
as first proposed by the Haldane model on a hexagonal lattice.\cite{haldane}
The quantized conductivity is given by $\sigma_{xy}={\cal C}(e^2/h)$
with the Planck constant $h$ and electronic charge $e$, and the Chern number
\ch~quantized due to topological restrictions.
\ch, whose origin involves occupied states entangled with conduction states, 
corresponds to the number of dissipationless and gapless edge states 
for the FM 2D Chern insulator.\cite{review}
The QAH insulator is anticipated to be a good candidate, with a great 
advantage in practice,
for anticipated applications of no energy consumption electronics\cite{review}, 
for Majorana fermions and their manipulation\cite{elliott},
and for future photonic devices.\cite{skirlo14}
High Chern number materials will provide
comparably higher conductivities.

A Chern insulating state arises in a broken time reversal system where spin-orbit
coupling (SOC) inverts valence and conduction bands which would otherwise provide
a trivial insulating phase. The topological gap is thereby limited by the strength
of SOC.  The interplay between strong interactions and strength
of SOC is being explored in the contexts of topologically insulating iridates\cite{iridate} and
possibly osmates,
but primarily model Hamiltonian treatments have explored (or suggested) the related phase diagram,
and none have followed how the phase transition occurs. Witczak-Krempa and
collaborators\cite{Witczak} have presented a heuristic phase diagram in which a
Chern insulating state borders a (trivial) Mott insulator.
However, modeling of the evolution of a realizable system through such a transition
is only now being addressed, with an example being the results of Doennig and co-workers\cite{rossitza15}
of the interplay between SOC and correlation effects in manipulating the competition
between Chern and Mott phases in a buckled (111) bilayer of LaFeO$_3$ in LaAlO$_3$.
Here we provide a related example for the bulk transition
metal oxide and Ising FM BaFe$_2$(PO$_4$)$_2$ (BFPO) whose structure is shown
in Fig. 1, of the competition between
SOC and strong interaction in creating and then annihilating a high Chern
number QAH phase.

The QAH phase has been predicted in various artificial structures
that can be roughly classified into three groups:
(1) topological insulators doped by magnetic transition metal (TM) 
ions\cite{fang08,fang10,fang14},
(2) thin TM layers on a hexagonal lattice\cite{zhang12,zhou14},
and (3) heterostructures of \{111\}-oriented double perovskite or 
TM oxides\cite{cook14,rossitza15,cai15,vand15}.  
A QAH state of \ch=$\pm2$ was calculated for
CrO$_2$/TiO$_2$\cite{cai15} and VO$_2$/TiO$_2$\cite{vand15}
heterostructures. However, having 
gaps of 2-4 meV leaves them primarily of academic interest.
A QAH state was also suggested and then calculated for
perovskite bilayers\cite{cook14,rossitza15},
but required a tuning of hybridization or trigonal distortion
to realize a QAH phase.
Zhang {\it et al.} suggested \ch=$\pm2$ QAH with a larger ($\sim 100$ meV) gap 
in graphene decorated with $5d$ transition-metal ions\cite{zhang12},
while Zhou {\it et al.} proposed the \ch=1 state with a larger gap around 100 meV
in a hexagonal tungsten lattice on the monolayer Cl-covered Si(111) surface\cite{zhou14}.
Generally, the QAH phase in these thin films has been expected only at a particular thickness.
Beyond the predicted systems, toy models\cite{wang11,tres12,yang12} 
with topological flat bands have suggested high Chern numbers 
in thin films or artificial heterostructures. A Chern insulating state in bulk solids remains
a goal for both research study and possible applications, and large Chern numbers
are especially valued.

\begin{figure*}[tp]
{\resizebox{16cm}{10cm}{\includegraphics{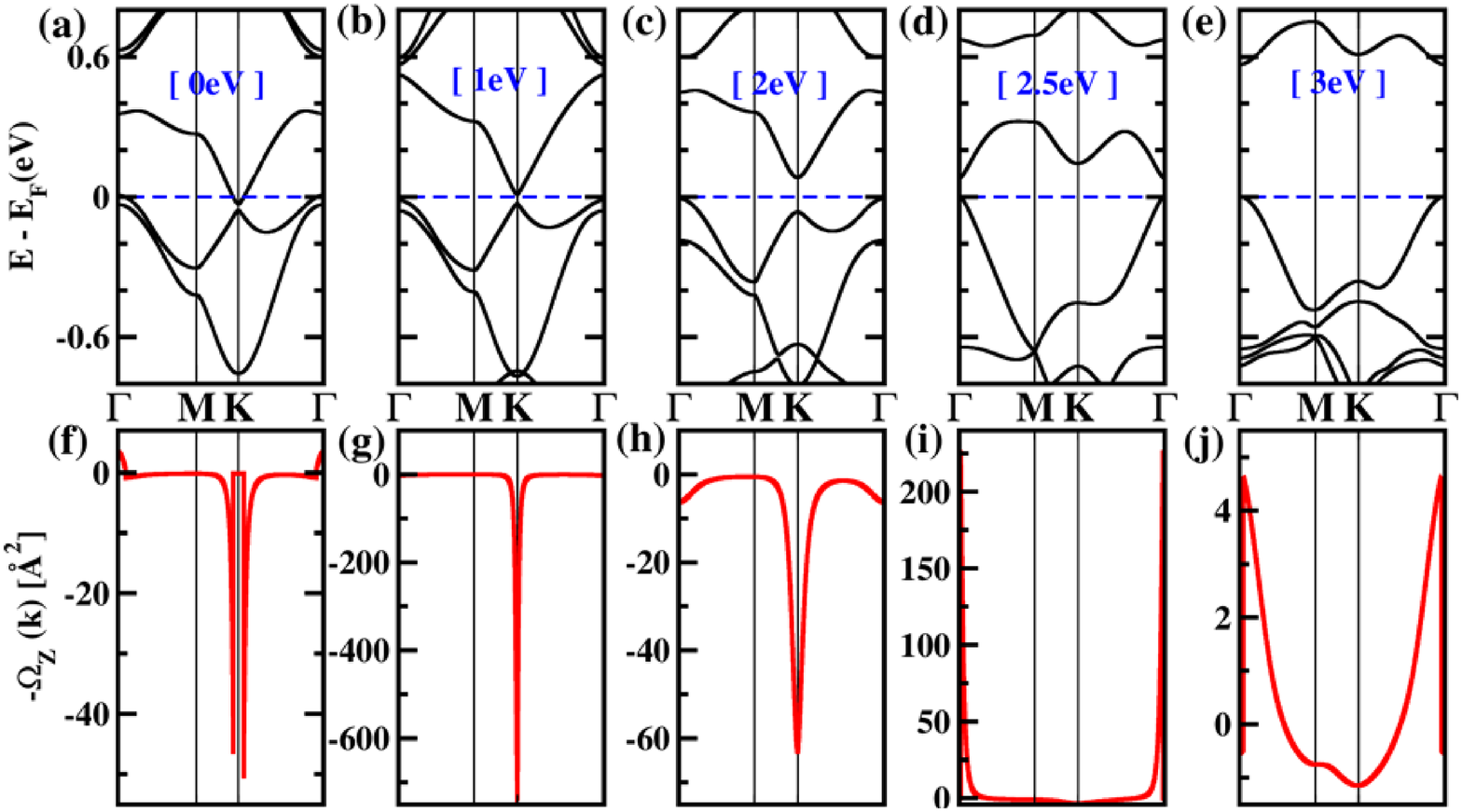}}}
\caption{(color online) Top row: 
 GGA+SOC+U band structures versus $U$, near the gap around 
 zero energy.  Two occupied minority bands lie below $E_F$=0. Bottom row:
 Berry curvatures $\Omega_k$ along
 the $\Gamma-M-K-\Gamma$ lines, also versus $U$.
 Note the different scales and signs of the vertical axes in
 the $\Omega_k$ plots. The origin of structure in $\Omega_k$ is
described in the text. 
}
\label{berry}
\end{figure*}

\begin{figure}[tp]
\vskip 8mm
{\resizebox{8.2cm}{4.8cm}{\includegraphics{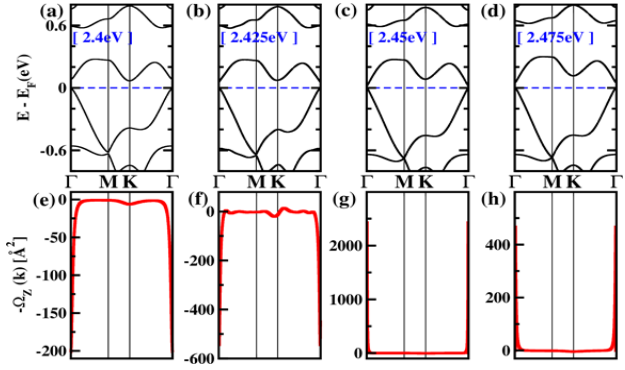}}}
\caption{(color online) As in Fig. 2, but for $U$ approaching 
$U_c$ on a fine scale.  The regime $U > U_c$ [panels (c) and (d)]
after band disentanglement
(which occurs at $\Gamma$, not at K), 
provides the non-topological phase in which the Mott gap increases from zero
and the orbital moment increases rapidly. Note
that the Berry curvature changes its predominant sign at $U_c$ (as well as
integrating to zero).
}
\label{berry2}
\end{figure}

Aside from these theoretically designed systems, 
a QAH realization has been reported
in the Cr-doped topological insulator (Bi,Sb)$_2$Te$_3$ with \ch=1
at very low temperature.\cite{chang}
To realize a QAH insulator with a large nontrivial gap, high Curie temperature $T_C$,
and high Chern number in real or easily synthesized compounds remains a challenge.
Morimoto and Nagaosa proposed,\cite{Nagaosa} based on a strong forward-scattering amplitude model, that
large gap topological insulators may be possible to achieve in strongly interacting systems, 
which shifts the focus away from $s-p$ materials.

In this paper we uncover, and then analyze the nature of, a transition between
gapped topological (Chern) and trivial (Mott) insulating states
in BFPO using a correlated band approach including SOC, by
tuning the interaction strength. 
With increasing interaction strength, a
compensated low-density Chern semimetal
evolves into a \ch=--3 Chern insulator with a gap up to 80 meV,
then transforms abruptly to a trivial Mott insulator, versus increasing strength of
on-site Coulomb repulsion $U$.
It is remarkable that the critical value for the transition is $U_{c}$ = 2.45 eV, a factor of
more than sixty greater than the SOC strength $\xi$=40 meV that produced the Chern phase. This
interplay between SOC and strong interaction highlights
how small energy scales can leverage topological restrictions to resist effects
of much larger energy scales.

\section{Structure, Symmetry, Methods}
\subsection{Structure and symmetry} 
Symmetries that are present can be critical for topological materials. 
Insulating BFPO\cite{bfpo15} crystallizes in space group $R{\bar 3}$ 
(threefold rotation plus inversion)
and is composed of layered 2D honeycomb sublattices of Fe$^{2+}$ ions within FeO$_6$ 
octahedra, as pictured in Fig. \ref{structure}. 
Due to a substantial interlayer separation of the Fe layers separated by Ba$^{2+}$ and 
(PO$_4$)$^{3-}$ insulating layers, BFPO is 2D electronically.
BFPO was synthesized by Mentr\'e and coworkers\cite{bfpo13,bfpo12},
who identified the rare 2D Ising FM nature with Curie temperature $T_C=65$ K. 
The very large calculated magnetocrystalline anisotropy is related to the
large orbital moment.
Our previous study\cite{bfpo15} provided the relaxed atomic positions
and revealed large exchange 
splitting $\Delta_{ex}$=3 eV of the $d^6$ ion,
enforcing the high spin $S=2$ configuration
with its filled and inactive majority $d$ orbitals. SOC will
couple the single minority $t_{2g}$ electron to majority orbitals 2-3 eV
removed in energy. Nevertheless, as we will see, SOC plays a critical
role in BFPO.

\subsection{Theoretical  methods}
The all-electron full-potential code {\sc wien2k}\cite{wien2k} 
incorporating density-functional-theory-based methods was applied,
with the structural parameters optimized in our previous study.\cite{bfpo15}
The Perdew-Burke-Ernzerhof generalized gradient approximation (GGA) 
was used as the exchange-correlation functional.\cite{gga}
The combined effects of correlation (Hubbard U) and SOC, needed to produce
the insulating state,  were included 
through GGA+U+SOC calculations.
Based on previous experience,\cite{bfpo15} $U$
is varied up to 4 eV, while the Hund's exchange parameter $J$ is fixed to 0.7 eV. 

From these results, a tight-binding Hamiltonian for the $t_{2g}$ manifold  
was obtained in terms of maximally localized Wannier functions\cite{marzari}
as implemented in {\sc wannier90}.
All necessary files for {\sc wannier90} were prepared by the 
code {\sc wien2wannier}.\cite{jan10}
The Brillouin zone (BZ), gapped almost everywhere, was sampled by $11\times11\times11$ $k$-mesh,
and 16 orbitals were used for the Wannier function projection.
Using {\sc wannier90},
the Berry curvature and the anomalous Hall conductivity were calculated 
with a very dense $k$-point grid of 5$\times$10$^5$ points to picture the Berry curvature 
and $300\times300\times300$ integration mesh to evaluate the anomalous Hall conductance. 


\section{Results and analysis}
\subsection{Band structure and Berry curvature}
First we  address the evolution of the band structure and Berry curvature
$\Omega_z({\vec k})$ versus increasing strength of $U$,
finding unanticipated behavior.
Although the perturbation theory expression for $\Omega_k$ is used in
the calculation, it is instructive to note the expression in terms of
the periodic part of the Bloch function
$u_k = w_k~exp(i\gamma_k)$ with non-negative magnitude $w_k$ and real
phase $\gamma_k$ (band indices suppressed),
\begin{eqnarray}
\Omega_k &=&-i\sum_{ij}\epsilon_{ij}<\partial_{k_i}u_k|\partial_{k_j}u_k>\nonumber \\
         &=& 2[<\partial_{k_x} w_k |w_k \partial_{k_y}\gamma_k>+
                 (x\leftrightarrow y)],
\label{berryphase}
\end{eqnarray}
where $\epsilon_{ij}$ is the rank-2 antisymmetric unit tensor.
The usual expression from perturbation theory suggests (as we find) that
regions with small bandgaps are important.\cite{vand15,vand06} These regions of small gap may
also involve larger matrix elements, because mixing by SOC is larger
when the gap is smaller.

Equation (\ref{berryphase}) makes explicit that a nonvanishing Berry curvature requires a k-dependent
phase. More specifically, it provides a different picture than the perturbation
theory expression (which is what is actually evaluated). The complementary interpretation
is that the Berry curvature obtains large contributions from where the gradients
of both $w_k$ and $\gamma_k$ are large. Evidently small gaps and large velocity
matrix elements promote such large gradients. In integer quantum Hall
effect systems the Chern number was related to zeros in $u_k$.\cite{Kohmoto}

\begin{figure}[tbp]
{\resizebox{7.5cm}{7.5cm}{\includegraphics{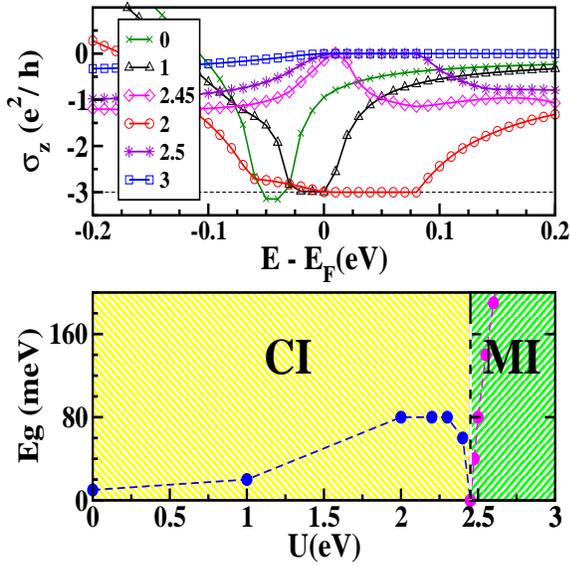}}}
\caption{(color online) Top panel: Anomalous Hall conductance $\sigma_{xy}$ 
in units of $e^2/h$ versus chemical potential, for the range of $U=0-3$ eV.
Below $U_c=2.45$ eV, a nontrivial gap $\sim$0.1 eV 
with Chern number of --3 appears, due to the spin-orbit driven gap
at the $K$ points.
Bottom panel: The minimum direct gap versus repulsion strength $U$. 
The Chern insulator (CI) phase
persists up to $U_c$; the minimum direct gap shifts from $K$ to $\Gamma$
around $U$=2.2 eV. 
Above $U_c$ lies the Mott insulator (MI) phase, where the gap increases rapidly with $U$  
to nearly 0.5 eV at $U$=4 eV.
} 
\label{ahc}
\end{figure}

\begin{figure}[tbp]
\vskip 6mm
{\resizebox{7.8cm}{4.5cm}{\includegraphics{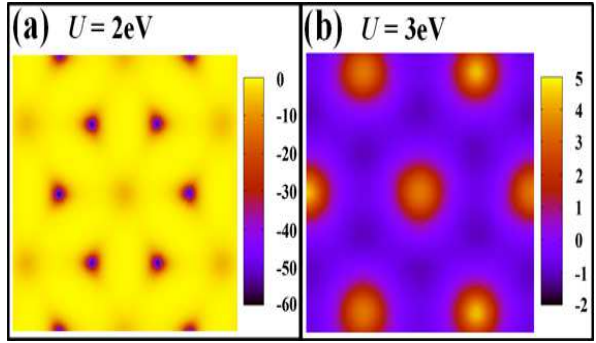}}}
\caption{(color online) Berry curvature $\Omega_k$ 
 in the entire zone. Left panel: $U$=2 eV, in the Chern phase (\ch=-3),
where strong peaks occur at the K points. Right panel:
$U$=3 eV, the \ch=0 trivial phase. Positive values of
$\Omega_k$ around the $\Gamma$ point are canceled by negative
values throughout the rest of the zone.
} 
\label{2d}
\end{figure}

In the uncorrelated limit $U~=~0$,
a tiny electron pocket in the Dirac point valley at $K$, visible in Fig. 2(a), 
is compensated with
a hole pocket at $\Gamma$,\cite{bfpo15} preventing the Dirac point from
pinning $E_F$.
SOC leads to the opening of a gap of 40 meV between the Dirac bands
at $K$, providing the SOC energy scale\cite{Bruno} $\xi$=40 meV.
The small gap results in sharp peaks appearing in Fig. \ref{berry}(f)
in the curvature $\Omega_z({\vec k})$ near the two $K$-points; the splitting into
two sharp peaks is caused by the Fermi level overlapping slightly the conduction band,
reproduced correctly by the fine k-point integration mesh.

\subsection{Evolution with interaction strength.}
For $U$ as small as 1 eV the band overlap disappears, the Dirac point pins the
Fermi energy as in graphene, and a sharp, almost  $\delta$-function like,
peak appears in $\Omega_z({\vec k})$ at K, shown in Fig. 2(g).
Unlike in graphene, here an orbital degree of freedom is involved. The Dirac
point degeneracy reflects the degeneracy in $R{\bar 3}$ symmetry
of the $e_g^{\prime}$ orbitals, the bands being
linear combinations of Fe orbitals with opposite orbital moment projections.
Incorporating SOC, these orbitals become entangled with the unbalanced spin
character, and a gap of several tens of meV
is opened at $K$ without inducing any significant orbital moment, {\it i.e.}
$e_g^{\prime}$ occupation is retained.

With further increase in $U$, the peaks in $\Omega_z({\vec k})$ 
broaden as the band structure evolves, with the gap at K increasing 
smoothly without significant orbital moment. 
For $U$ up to a critical value $U_c$, SOC retains its hold on the Chern
state (see below), in spite of the large value of $U$ and of spin-exchange splitting
$\Delta_{ex}$, compared to $\xi$. The bands and Berry curvature calculated
for values of $U$ from 2.40 eV to 2.475 eV are shown in Fig. 3.
Above $U$=2 eV the gap between valence and conduction bands at $\Gamma$
decreases rapidly, closing at the critical value $U_c$=2.45 eV as
shown in Fig. 4, lower panel. This
gap closing and reopening marks a disentanglement of bands and loss
of topological character, as we verify below. Note that the eigenvalue
crossing at $\Gamma$ does not lead to a metallic trivial state, but
to a Mott insulating state dictated by the Coulomb repulsion $U$.
The vanishing of the Chern number at the transition suggests that the character 
of the Wannier functions of the two filled bands will have
changed discontinuously at the critical point. The character of the
Berry curvature changes dramatically, as shown in Fig. 5.

\begin{figure}[tbp]
{\resizebox{7.6cm}{6cm}{\includegraphics{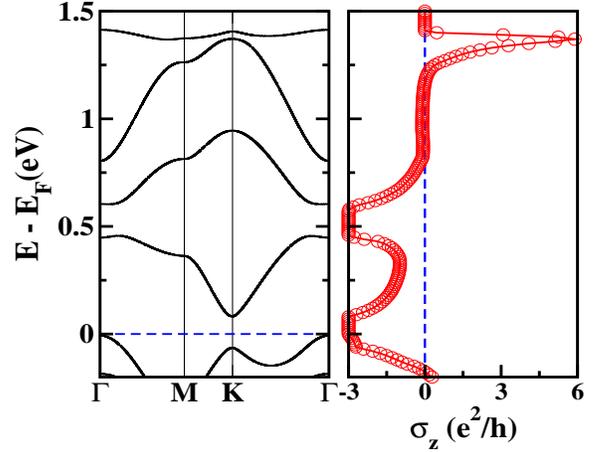}}}
\caption{(color online) Enlarged GGA+SOC+U band structures (left) and
energy resolved
anomalous Hall conductance $\sigma_{xy}$ (right) in units of $e^2/h$ at $U=2$ eV.
The two bands below the gap (horizontal line; one band is not visible)
provide \ch=-3; the next
higher band is non-topological $\Delta$\ch=0; entanglement of the top band
with the next lower one gives it a very large Chern number \ch=6.
} 
\label{ahc2ev}
\end{figure}

\begin{figure*}[tbp]
{\resizebox{15cm}{10.3cm}{\includegraphics{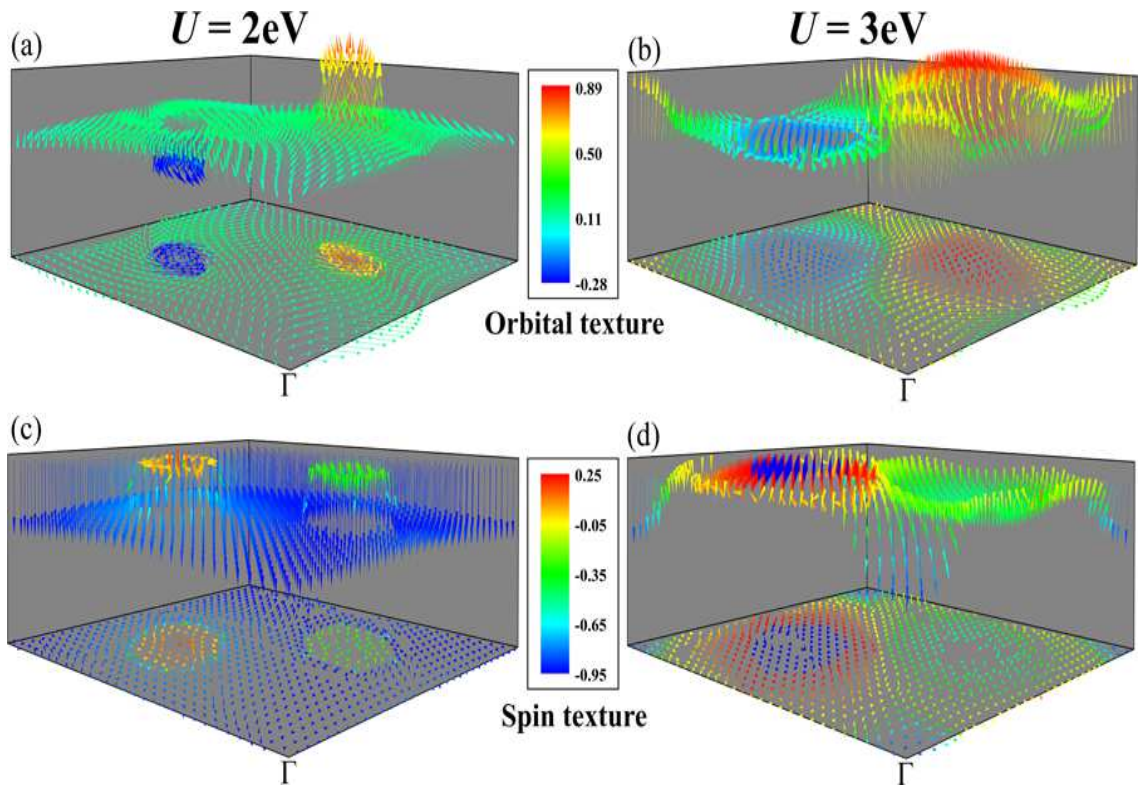}}}
\caption{(color online)
Orbital $\sum_n\langle u_{kn}|\vec{l}|u_{kn}\rangle$ 
and spin $\sum_n\langle u_{kn}|\vec{s}|u_{kn}\rangle$
textures in a square region in $\vec k$ space. The axes are
along Cartesian axes; $\Gamma$ is at the near corner, $M$ is midway
along the diagonal toward the far corner, the vortices circle the
K, K' points. (a) The orbital texture in the Chern phase
($U$=2 eV) and (b) in the $U$=3 eV trivial Mott insulator. (c), (d)
The corresponding spin textures.
The color denotes the $\hat z$  component of the texture field,
with positive being parallel to the spin orientation, {\it i.e.}, $\hat{c}$-direction,
while the arrow provides the direction. The bottom plane provides the
$\hat z$ projection  to allow better visualization of the in-plane
variation.
} 
\label{orb}
\end{figure*}

We had previously found\cite{bfpo15} that beyond $U\approx$ 2.5 eV the orbital
moment $m_{\ell}$ increases rapidly from $\sim$0.1$\mu_B$ to 0.5--0.6$\mu_B$,
asymptoting to the remarkably large value (for a $3d$ ion) of 0.7$\mu_B$
for $U>$ 5 eV. 
This evolution reflects a reoccupation of orbitals. In the Chern phase
the minority $t_{2g}$ electron is in a linear combination with almost equal
amounts of orbital component $m_{\ell}=\pm 1$ and perhaps some 
$m_{\ell}=0$, while beyond $U_c$ the $m_{\ell}=+1$ orbital becomes
fully occupied.
The spin remains close to its S=2 value.
This jump in orbital moment signals the reoccupation from a linear
combination of the two $e_g^{\prime}$ orbitals to dominant occupation of the
complex $m_{\ell}=+1$ linear combination, a sign that SOC retains influence
even after the loss of topological character. The topological-to-trivial
transition thus results from a continuous, strong interaction driven
band reordering,
manipulated by the interplay between $U$ and the SOC
energy $\xi$=40 meV. The transition occurs at a ratio $U/\xi$=60.

\subsection{Chern number and its origin}
A chemical potential resolved Chern number \ch(E) is obtained 
from the anomalous Hall conductivity
$\sigma_{xy}$ = \ch~$e^2/h$ by integrating the Berry curvature
$\Omega_k$ over the zone, up to the chemical potential.
For values of $U$ increasing through $U_c$,
the energy-resolved $\sigma_{xy}(E)$ near $E_F$ is displayed in 
the top panel of Fig.~\ref{ahc}. The bottom panel of this figure
shows the evolution of the gap, increasing from zero at $U$=0 and
achieving 80 meV at $U$=2 eV. In the range $U$=2.0-2.4 eV the minimum
gap moves from its position at $K$ to $\Gamma$.
In this range BFPO is a Chern insulator with \ch=~-3, signaling a
quantum anomalous Hall
phase with {\it three times} the minimum quantized anomalous Hall conductivity.
Rapidly with $U$ increasing from 2.40 eV, the gap at $\Gamma$ closes and
re-opens due to the interchange in energy of the two bands at $\Gamma$.
At $U=2.5$ eV, the contribution of the very sharp  positive peak at
$\Gamma$
is canceled by smaller negative contributions elsewhere, and  \ch$\rightarrow$0;
BFPO has transformed to a trivial insulator.

Besides the fundamental gap, there are additional nontrivial gaps at higher energies.
The right panel of Fig. \ref{ahc2ev} shows $\sigma_{xy}$ in the
--0.25 eV to 1.5 eV range for $U=2$ eV.
Overall, there are three nontrivial gaps, about 80 meV wide at $E_F$ and 100 meV at 0.5 eV, 
and a smaller gap at 1.3 eV.
Notably, the highest band has \ch=--6, while the two bands just below provide
\ch=+9 to the running sum. 
Strong spin-orbit entanglement extends throughout the $t_{2g}$ bands in
the Chern phase.
The nontrivial states can be attributed to topological
spin-orbit band entanglement, at the K points for
the physical, fundamental gap;
at the $\Gamma$ point for band filling up to 0.5 eV; again at the 
K points for band filling to 1.3 eV
(see the left panel of Fig. \ref{ahc2ev}). 
Above $U_c$, the last two gaps increase a little, whereas the lowest one
becomes trivial.

\section{Angular momenta and their textures}
SOC produces noncollinear texture of the spin and orbital magnetizations,
as studied early on in real space by Nordstr\"om and 
Singh.\cite{noncollinear} Here we
study the spin and orbital moment texture in k-space, comparing its behavior
in the Chern insulator phase to that in the Mott insulator phase. The
results indicate effects of spin-orbit coupling can be strongly
confined to specific regions of the zone, but in a manner very different
from the high but narrow peaks of the Berry curvature.

The contribution to the orbital moment (and analogously for the spin
moment) at point $\vec r$ from point $\vec k$ in the zone is
\begin{eqnarray}
m^{\ell}_k(\vec r)=\sum_n^{occ}u_{k,n}^*(\vec r)\vec{l}u_{k,n}(\vec r),
\end{eqnarray} 
in terms of the two occupied minority band wavefunctions 
$u_{k,1}$ and $u_{k,2}$.
These quantities can be summed over the zone to give the texture
in real space, or integrated over the cell to give the texture in
the zone. It is this latter approach that we address here, since
topological character is connected directly to the $\vec k$-dependence 
of the Bloch functions.

With the hope of uncovering additional aspects of the nature of the 
topological transition, we present in Fig.~\ref{orb}
the orbital moment $\sum_n \langle u_{k,n}|\vec{l}|u_{k,n}\rangle$ 
and spin $\sum_n \langle u_{k,n}|\vec{s}|u_{k,n}\rangle$ textures
summed over the two occupied minority bands, 
which are the orbitally active ones in the $t_{2g}$ manifold.
The band decomposed analogs are presented in the Supplementary Material.\cite{supp}
One can readily see that the texture is highly structured, compared
for example to the simple vortex shape found on Cu(111) and Au(111)
surfaces by Kim {\it et al.}\cite{Kim2012}

In the Chern phase $U<U_c$ (left panels of Fig.~\ref{orb})
both orbital and spin textures are slowly varying
near their mean values except for sharply defined elliptical regions
around the K and K' points, where a significant disruption appears. 
Most notably, the orbital texture displays
chiral character of oppositely oriented circulations, reminiscent of the
source-sink character found in the Berry connection in honeycomb lattice models.
The spin is large everywhere, consistent with the ferromagnetic character,
with texture that also displays well defined,
but different size and shape, regions around the K and K'
points. The $z$-components differ in the two regions, both being
substantially different from the mean.

Beyond the critical value $U_{c}$, the Mott insulating state
appears with enhanced $m_{\ell}=+1$ character throughout the zone.
There is a small orbital moment $\sim$0.1$\mu_B$ for $U < U_c$
but it increases rapidly at larger $U$.
The corresponding textures are shown in the right panels of Fig. \ref{orb}.
The transition from Chern to Mott insulator at $U_c$
is accompanied by a change in character of both orbital and spin texture fields.
Now the effect of SOC in coupling spin texture to orbital texture
dies rather abruptly away from the K, K' points, and structure has appeared
around $\Gamma$.

It should be mentioned that, unlike the Chern number, 
the change between the textures at $U$=2 eV and
$U$=3 eV is, from its definition, not abrupt. The value of the (say) orbital field
$\sum_n \langle u_{k,n}|\vec{l}|u_{k,n}\rangle$ at one point $\vec k$ does
not know what is occurring at other k points, nor does it know specifically
about global topological properties; it varies continuously with changes,
including changes in $U$, as long as no first-order transition is
encountered. 

Some general observations can be made. A strictly local isotropic atomic moment should 
correspond to a smooth, symmetric texture in k-space. This is not observed,
so the Fe moment has substantial itinerant character.  A uniform (purely
itinerant) magnetization in real space would arise from a strong peak 
at small $|k|$. This also is not observed, so the Fe moment does have substantial
local character. It might also be relevant that the Fe moment is calculated to
have extremely large magnetocrystalline anisotropy. The sharp
boundaries of regions differing from featureless texture are much more
prominent in the Chern phase than in the trivial phase, indicating that
real space texture evolves in both the orbital and spin moments as the
orbital moment becomes larger and better defined. 

\section{Summary}
In this work we have followed the evolution of the honeycomb lattice Ising
ferromagnet  BaFe$_2$(PO$_4$)$_2$  from Chern insulator at small to moderate 
interaction strength U to the Mott insulator
phase beyond the critical strength $U_c$=2.45 eV, the latter being the physical regime. 
A noteworthy aspect is that the Chern phase has \ch = --3 and, at its maximum, a
sizable gap of 80 meV. For small $U$ after SOC is included, the Chern
insulator phase is obtained within which
the gap at K increases very slowly with $U$. The small SOC strength $\xi$=40 meV
is sufficient to withstand increasing $U$ and support the topological phase
 up to the critical value $U_c$=2.45 eV. This support is possible because the
increasing (with $U$) gap at $K$ is supplanted by a decreasing gap at $\Gamma$.
However, the gap at $\Gamma$ closes rapidly,
and immediately re-opens with the 
system in a trivial Mott insulating phase (\ch = 0).
For Fe in an oxide insulator, $U$ will be at least 4--5 eV in magnitude, so
we do not expect BaFe$_2$(PO$_4$)$_2$ to show Chern insulating properties. 
Evidently a larger strength of SOC or smaller $U$ is the direction to search
for a Chern insulator in this class.

For  additional insight,
we have demonstrated that the spin and
angular momentum textures throughout the zone experience  an evolution from a tight
structure around the K and K' points in the Chern phase to a more extended 
character in the Mott phase, varying more
smoothly throughout
the zone.  

A distinctive feature of BFPO is the large Chern number \ch = --3. In fact, the
highest lying minority $t_{2g}$ band has \ch = --6, with the intermediate bands
contributing \ch = +9. Ren {\it et al.}\cite{Ren} in their review have commented
on large Chern number systems, and referenced the few that have been predicted,
mostly in model systems. As an example, Jiang {\it et al.}\cite{Jiang}
have calculated Chern numbers for multilayer films in magnetic fields for which
the Chern number, for tuned field values, can be as large as the number of layers
(up to 12 in their model).

\begin{acknowledgments}
 We acknowledge A. S. Botana for technical discussions,
 A. Essin and K. Park for conversations on topological insulators, J. Kune\v{s}
 for assistance on angular momentum texture, and
 R. Pentcheva for discussions of related behavior in buckled 
 (111) transition metal oxide bilayers.
 This research was supported by National Research Foundation of Korea
 Grant No. NRF-2013R1A1A2A10008946 and NRF-2016R1A2B4009579 at Korea University,
 and by US Department of Energy Grant NO. DE-FG02-04ER46111 (W.E.P.).
\end{acknowledgments}


\begin{thebibliography}{10}
\bibitem{haldane} F. D. M. Haldane,
Model for a Quantum Hall Effect without Landau Levels: 
Condensed-Matter Realization of the ``Parity Anomaly",
Phys. Rev. Lett. {\bf 61}, 2015 (1988).

\bibitem{review} For a review, see H. Weng, R. Yu, X. Hu, X. Dai, and Z. Fang,
 Quantum anomalous Hall effect and related topological electronic states,
 Adv.  Phys. {\bf 64}, 227 (2015).

\bibitem{elliott} S. R. Elliott and M. Franz,
 Majorana fermions in nuclear, particle, and solid-state physics,
 Rev. Mod. Phys. {\bf 87}, 137 (2015).


\bibitem{skirlo14} S. A. Skirlo, L. Lu, and M. Solja\v{c}i\'c,
 Multimode one-way waveguides of large Chern numbers,
  Phys. Rev. Lett. {\bf 113}, 113904 (2014).

\bibitem{iridate}J.-M. Carter, V. V. Shankar, M. A. Zeb, and H.-Y. Kee,
Semimetal and Topological Insulator in Perovskite Iridates,
Phys. Rev. B {\bf 85}, 115105 (2012).

\bibitem{Witczak}W. Witczak-Krempa, G. Chen, Y. B. Kim, and L. Balents,
 Correlated quantum phenomena in the strong spin-orbit regime,
 Annu. Rev. Condens. Matter Phys. {\bf 5}, 57 (2014).

\bibitem{rossitza15} D. Doennig, S. Baidya, W. E. Pickett, and R. Pentcheva,
Design of Chern and  Mott insulators in buckled $3d$-oxide honeycomb
lattices,
Phys. Rev. B {\bf 93}, 165145 (2016).

\bibitem{fang08} C.-X. Liu, X.-L. Qi, X. Dai, Z. Fang, and S.-C. Zhang,
Quantum anomalous Hall effect in Hg$_{1-y}$Mn$_y$Te quantum wells,
Phys. Rev. Lett. {\bf 101}, 146802 (2008).

\bibitem{fang10} R. Yu, W. Zhang, H.-J. Zhang, S.-C. Zhang, X. Dai, and Z. Fang,
 Quantized anomalous Hall effect in magnetic topological insulators,
 Science {\bf 329}, 61 (2010).

\bibitem{fang14} C. Fang, M. J. Gilbert, and B. A. Bernevig,
Large-Chern-number quantum anomalous Hall effect in thin-film 
topological crystalline insulators,
Phys. Rev. Lett. {\bf 112}, 046801 (2014). 

\bibitem{zhang12} H. Zhang, C. Lazo, S. Bl\"ugel, S. Heinze,
 and Y. Mokrousov,
Electrically tunable quantum anomalous Hall effect in graphene decorated
by $5d$ transition-metal adatoms, 
Phys. Rev. Lett. {\bf 108}, 056802 (2012).

\bibitem{zhou14} M. Zhou, Z. Liu, W. Ming, Z. Wang, and F. Liu,
$sd^2$ graphene: Kagome band in a hexagonal lattice,
Phys. Rev. Lett. {\bf 113}, 236802 (2014).

\bibitem{cook14} A. M. Cook and A. Paramekanti,
Double Perovskite heterostructures: magnetism, Chern bands, 
and Chern insulators,
Phys. Rev. Lett. {\bf 113}, 077203 (2014).


\bibitem{cai15} T. Cai, X. Li, F. Wang, S. Ju, J. Feng, and C.-D. Gong,
Single-spin Dirac Fermion and Chern insulator based on simple oxides,
Nano Lett. {\bf 15}, 6434 (2015).

\bibitem{vand15} H. Huang, Z. Liu, H. Zhang, W. Duan, and D. Vanderbilt,
 Emergence of a Chern-insulating state from a semi-Dirac dispersion,
 Phys. Rev. B {\bf 92}, 161115(R) (2015).

\bibitem{wang11} F. Wang and Y. Ran,
 Nearly flat band with Chern number $C=2$ on the dice lattice,
Phys. Rev. B {\bf 84}, 241103(R) (2011).

\bibitem{tres12} M. Trescher and E. J. Bergholtz,
Flat bands with higher Chern number in pyrochlore slabs,
Phys. Rev. B {\bf 86}, 241111(R) (2012).

\bibitem{yang12} S. Yang, Z.-C. Gu, K. Sun, and S. Das Sarma,
Topological flat band models with arbitrary Chern numbers,
Phys. Rev. B {\bf 86}, 241112(R) (2012).


\bibitem{chang} C.-Z. Chang, J. Zhang, X. Feng, J. Shen, Z. Zhang, M. Guo, 
 K. Li, Y. Ou, P. Wei, L.-L. Wang, Z.-Q. Ji, Y. Feng, S. Ji, X. Chen, J. Jia,
 X. Dai, Z. Fang, S.-C. Zhang, K. He, Y. Wang, L. Lu, X.-C. Ma,
 and Q.-K. Xue,
 Experimental Observation of the Quantum Anomalous Hall Effect 
 in a Magnetic Topological Insulator,
 Science {\bf 340}, 167 (2013).


\bibitem{Nagaosa}T. Morimoto and N. Nagaosa,
  Weyl Mott Insulator, Sci. Rep. {\bf 6}, 19853 (2016).
doi:10.1038/srep19853


\bibitem{bfpo15} Y.-J. Song, K.-W. Lee and W. E. Pickett,
 Large orbital moment and spin-orbit enabled Mott transition 
 in the Ising Fe honeycomb lattice \bfpo,
 Phys. Rev. B {\bf 92}, 125109 (2015).


\bibitem{bfpo13} R. David, A. Pautrat, D. Filimonov, H. Kabbour, H. Vezin, 
 M.-H. Whangbo, and O. Mentr\'e,
 Across the structural re-entrant transition in \bfpo: 
 Influence of the two-dimensional ferromagnetism, 
 J. Am. Chem. Soc. {\bf 135}, 13023 (2013).

\bibitem{bfpo12} H. Kabbour, R. David, A. Pautrat, H.-J. Koo, 
M.-H. Whangbo, G. Andr\'e, and Mentr\'e,
A genuine twodimensional Ising ferromagnet with magnetically 
driven re-entrant transition.
Angew. Chem. Int. Ed. {\bf 51}, 11745 (2012).

\bibitem{wien2k} K. Schwarz and P. Blaha,
 Solid state calculations using WIEN2k,
 Comput. Mater. Sci. {\bf 28}, 259 (2003).

\bibitem{gga} J. P. Perdew, K. Burke, and M. Ernzerhof, 
 Generalized gradient approximation made simple, 
 Phys. Rev. Lett. {\bf 77}, 3865 (1996).


\bibitem{marzari} A. A. Mostofi, J. R. Yates, Y.-S. Lee, I. Souza, D. Vanderbilt,
 and N. Marzari,
 wannier90: A tool for obtaining maximally-localised Wannier functions,
 Comput. Phys. Commun.{\bf 178}, 685 (2008).

\bibitem{jan10} J. Kune\v{s}, R. Arita, P. Wissgott, A. Toschi,
 H. Ikeda, and K. Held,
 Wien2wannier: From linearized augmented plane waves to maximally 
 localized Wannier functions,
 Comput. Phys. Commun. {\bf 181}, 1888 (2010).


\bibitem{vand06} X. Wang, J. R. Yates, I. Souza, and D. Vanderbilt,
 {\it Ab initio} calculation of the anomalous Hall conductivity 
 by Wannier interpolation,
 Phys. Rev. B {\bf74}, 195118 (2006).

\bibitem{Kohmoto} M. Kohmoto,
  Topological Invariant and the Quantization of the Hall Conductance,
  Ann. Phys. (N.Y.) {\bf 160}, 343 (1985).

\bibitem{Bruno} The SOC coefficient of the $3d$
shell of atomic Fe is 80 meV, see
O. \v{S}ipr, M. Ko\v{s}uth, and H. Ebert,
Magnetic structure of free iron clusters compared to iron crystal surfaces,
  Phys. Rev. B {\bf 70}, 174423 (2004). The bands in BFPO are hybridized
with O $2p$ states, decreasing the effective SOC strength to 40 meV.

\bibitem{noncollinear}L. Nordstr\"om and D. J. Singh,
 Noncollinear intra-atomic magnetism,
 Phys. Rev. Lett. {\bf 76}, 4420  (1996).

\bibitem{supp} See Supplementary Material for the band-resolved spin 
and orbital textures, which show similar behavior as in Fig. \ref{orb}.

\bibitem{Kim2012}B. Kim, C. H. Kim, P. Kim, W. Jung, Y. Kim, Y. Koh,
M. Arita, K. Shimada, H. Namatame, M. Taniguchi, J. Yu, and C. Kim,
Spin and orbital angular momentum structure of Cu(111) and Au(111)
surface states,
 Phys. Rev. B {\bf 85}, 195402 (2012).

\bibitem{Ren} Y. Ren, Z. Qiao, and Q. Niu,
Topological Phases in Two-Dimensional Materials: a Brief Review,
Rep. Prog. Phys. {\bf 79}, 066501 (2016).

\bibitem{Jiang}H. Jiang, Z. Qiao, H. Liu, and Q. Niu,
 Quantum anomalous Hall effect with tunable Chern number in
magnetic topological insulator film,
Phys. Rev. B {\bf 85}, 045445 (2012).


\end{thebibliography}
\end{document}